\documentclass[conference,a4paper]{IEEEtran}

\usepackage{amsmath,epsfig,amssymb,verbatim}
\usepackage[nocompress]{cite}
\usepackage{graphicx}
\usepackage{bm,bbm}
\usepackage{xcolor}
\usepackage{subfigure}
\usepackage{stackengine}
\usepackage{cleveref}
\usepackage{multicol}
\usepackage{multirow}

\def\@IEEEBIOskipN{1\baselineskip}

\newtheorem{Lemma}{Lemma}
\newtheorem{Proposition}{Proposition}

\begin{document}
\title{Artificial-Noise Aided Design for Movable-Antenna Enabled Physical-Layer Service Integration}

\author{\IEEEauthorblockN{Zhifeng Tang$^{\ast}$, Guangchen Wang$^{\ast}$, Nan Yang$^{\ast}$, Xiangyun Zhou$^{\ast}$, and Salman Durrani$^{\ast}$}
\IEEEauthorblockA{$^{\ast}$School of Engineering, Australian National University, Canberra, ACT 2600, Australia}
\IEEEauthorblockA{Emails: \{zhifeng.tang, guangchen.wang, nan.yang, xiangyun.zhou, salman.durrani\}@anu.edu.au}
}

\maketitle

\begin{abstract}
This paper pioneers a novel scheme for artificial-noise (AN)-aided movable-antenna (MA)-enabled physical-layer service integration (PLSI) to harmonize the simultaneous delivery of multicast and confidential messages. By jointly exploiting the spatial reconfiguration capability of MAs and the interference shaping capability of AN, we aim to enhance secrecy performance while guaranteeing multicast reliability. The joint design of MA positions and transmit variables results in a highly coupled and non-convex optimization problem. To address this, we first provide key insights into the role of spatial degrees of freedom in AN design. We then characterize the AN direction under a structured transmission design and derive a closed-form expression for the AN-to-confidential power allocation ratio, which significantly simplifies the overall design. To solve the resulting problem, we further develop a low-complexity block coordinate ascent (BCA)-based scheme that alternates between transmit design and MA position optimization. Numerical results demonstrate that the proposed scheme achieves significant secrecy performance gains with low computational complexity and fast convergence, highlighting its effectiveness for MA-enabled PLSI systems.
\end{abstract}


\section{Introduction}

The rapid evolution of next-generation wireless networks, particularly in the sixth-generation (6G) systems, calls for highly flexible and adaptive transmission architectures \cite{Zhang2019vtm,Tang2023JSAC}. Conventional fixed-position antenna arrays are inherently limited in their ability to dynamically adapt to channel variations and spatial user distributions, especially in highly dynamic and complex propagation environments \cite{Ning2025wcom,Zhu2026tcst}. To address this, the concept of movable antennas (MAs) has recently emerged as a promising paradigm, where antenna elements can be physically repositioned within a predefined region to better match the propagation environment \cite{Zhu2024twc}. By enabling spatial reconfiguration of antenna positions, MA unlocks new spatial degrees of freedom (DoFs), which allows for the reshaping of propagation environments to improve transmission performance, a potential already substantiated by recent hardware prototypes and experimental demonstrations \cite{Dong2026twc}.

Harnessing the benefits of spatial reconfiguration, recent literature has systematically investigated the fundamental principles and developed MA-enabled transmission, demonstrating significant performance gains across various communication scenarios \cite{Xiao2024twc,Zhang2025tvt,Irshad2025wcoml}. This architectural agility also facilitates the application of MAs in physical-layer security, where its effectiveness in enhancing secrecy performance has been demonstrated. Initial efforts focused on single-user scenarios, where antenna positions are optimized to improve secrecy rate by enhancing the legitimate link while suppressing the eavesdropper's link \cite{Hu2024spl}. These designs subsequently evolved into multi-user scenarios, such as non-orthogonal multiple access (NOMA) systems featuring multiple legitimate users and eavesdroppers \cite{Mao2025wcl}. More recent studies further incorporated artificial noise (AN) to actively interfere with eavesdroppers \cite{Ding2025twc,Tang2025tcom}, as well as considered more practical scenarios with imperfect or unknown eavesdropper's channel state information (CSI) \cite{Hu2024tmc}. However, these advancements remain tethered to conventional secure communication scenarios, leaving a critical void in scenarios requiring the coexistence of multiple services with heterogeneous requirements.

To support such heterogeneous service requirements, future wireless systems are gravitating towards unified transmission frameworks capable of the concurrent delivery of public multicast and confidential services \cite{Schaefer2014spm}. These frameworks, formalized as physical-layer service integration (PLSI) \cite{Ning2021twc}, necessitate a delicate equilibrium: the transmitter needs to ensure reliable reception of the multicast message for all users while preventing unintended users from decoding the confidential message.
This dual-service requirement triggers a fundamental tradeoff between multicast reliability and secrecy performance, as the shared power resources must be strategically allocated between multicast and confidential messages. Recent few studies have shown that incorporating MAs into PLSI systems provides additional spatial flexibility to improve overall system performance \cite{Shen2025wcl}. Nevertheless, whether and how AN can be effectively integrated with MA-enabled spatial reconfiguration to further enhance secrecy performance while preserving multicast reliability presents an intricate design challenge, which motivates this work.



\subsubsection*{Our Contributions} In this work, we propose an innovative joint design for AN-aided MA-enabled PLSI systems to enhance secrecy performance while ensuring multicast reliability, where we exploit the spatial reconfiguration capability of MAs and the interference shaping capability of AN. Particularly, we develop a low-complexity transmission scheme by characterizing the AN direction within a structured framework and deriving a closed-form expression for the AN-to-confidential power allocation ratio, which significantly simplifies the overall design. Building on this, we propose a block coordinate ascent (BCA)-based scheme to jointly optimize MA positions and transmit variables. Numerical results demonstrate that our proposed scheme provides a substantial leap in the secrecy rate performance over conventional fixed-antenna and MA-enabled benchmark schemes, while maintaining fast convergence and low complexity, establishing it as a highly practical solution for next-generation secure service integration.


\section{System Model}\label{Sec:System}


We consider the downlink of a wireless communication system 
where a base station (BS), equipped with a MA platform containing $M$ MAs, serves two single-antenna user equipments (UEs), denoted by $U_1$ and $U_2$. In this system, the BS simultaneously transmits a confidential message intended for $U_1$ and a multicast message intended for both UEs. Although $U_2$ is a legitimate UE for the multicast message, it may attempt to decode the confidential message for $U_1$, and thus acts as a potential eavesdropper.

We establish a three-dimensional (3D) Cartesian coordinate system where the BS is located at the origin. The antenna aperture lies on the $x$-$z$ plane, with the $y$-axis perpendicular to the aperture. Due to mechanical constraints, each MA is restricted to move within the $x$-$z$ plane. Accordingly, the position of the $m$-th MA is denoted by $\mathbf{q}_m = [x_m, 0, z_m]^T$, where $x_m$ and $z_m$ denote the horizontal and vertical coordinates of the antenna within the aperture, respectively, and $(\cdot)^{T}$ is the transpose. We assume that each MA is confined to a common predefined aperture region $\mathcal{Q}$, i.e., $\mathbf{q}_m \in \mathcal{Q}$, $\forall m$. The positions of all MAs are collected in the matrix $\mathbf{Q} =[\mathbf{q}_1,\mathbf{q}_2,\ldots,\mathbf{q}_M]$,
where $\mathbf{Q}\in\mathbb{R}^{3\times M}$ and each column corresponds to the position of one MA. To avoid antenna collision and excessive electromagnetic coupling, a minimum spacing constraint between any two antennas is imposed as $\|\mathbf{q}_i-\mathbf{q}_j\| \ge d_{\min}$, for all $i \neq j$, where $d_{\min}$ denotes the minimum inter-MA distance and $\|\cdot\|$ is the Euclidean norm.

\subsection{Channel Model}

We adopt a geometric multipath channel model, where the BS-UE link consists of multiple scattering paths. We denote $\mathbf q_0 = [0,0,0]^T$ as the reference position of the MA. Under the plane-wave approximation, the channel coefficient between the $m$-th MA and $U_k$ is expressed as
\begin{equation}
h_{k,m}(\mathbf q_m)=\sum_{\ell=1}^{L_k}\alpha_{k,\ell}e^{-j\frac{2\pi}{\lambda}\mathbf u_{k,\ell}^T(\mathbf q_m-\mathbf q_0)},
\end{equation}
where $L_k$ denotes the number of propagation paths between the BS and $U_k$, and $\alpha_{k,\ell}$ represents the complex gain of the $\ell$-th path at the reference position $\mathbf q_0$, capturing both large-scale attenuation and small-scale fading \cite{Zhu2024twc}. Here, $\lambda = \frac{c}{f_c}$ denotes the carrier wavelength, where $c$ is the speed of light and $f_c$ is the carrier frequency, and $\mathbf u_{k,\ell}\in\mathbb{R}^{3\times1}$ is the unit propagation direction vector of the $\ell$-th path. Specifically, $\mathbf u_{k,\ell}$ is given by
\begin{align}
\mathbf u_{k,\ell}=
[\sin\theta_{k,\ell}\cos\phi_{k,\ell},
\cos\theta_{k,\ell},
\sin\theta_{k,\ell}\sin\phi_{k,\ell}]^T,
\end{align}
where $\theta_{k,\ell}$ is the elevation angle measured from the $y$-axis and $\phi_{k,\ell}$ is the azimuth angle in the $x$-$z$ plane. Since both UEs are legitimate, we assume perfect instantaneous CSI of all links to be available at the BS, i.e., the parameters $\{L_k,\alpha_{k,\ell},\mathbf u_{k,\ell}\}$ are perfectly known.

Since all MAs are confined within a small aperture region $\mathcal{Q}$ at the BS, they share the same multipath parameters $\{L_k,\alpha_{k,\ell},\mathbf u_{k,\ell}\}$, while their distinct spatial positions $\{\mathbf q_m\}$ introduce path-dependent phase shifts. By stacking the channel coefficients associated with all $M$ antennas, the channel vector between the BS and $U_k$ is given by
\begin{equation}
\mathbf h_k(\mathbf Q)=[h_{k,1}(\mathbf q_1),\ldots,h_{k,M}(\mathbf q_M)]^T \in \mathbb{C}^{M\times1}.
\end{equation}

\subsection{Transmit Signal and Secrecy Rate}

The BS employs linear precoding to transmit a confidential message intended for $U_1$ and a multicast message intended for both UEs. We denote $\mathbf{x}\in \mathbb{C}^{M\times 1}$ as the signal vector transmitted from the BS to both UEs, given by $\mathbf{x} = \mathbf{w}_1 s_1 +\mathbf{w}_0 s_0+\mathbf{v}z$, where $\mathbf{w}_1\in \mathbb{C}^{M\times1}$ and $\mathbf{w}_0 \in \mathbb{C}^{M\times1}$ denote the beamforming vectors for the confidential and multicast signals, respectively, $s_1$ denotes the confidential signal for $U_1$, $s_0$ denotes the multicast signal intended for both UEs, $\mathbf{v}\in\mathbb{C}^{M\times1}$ denotes the AN vector, and $z$ denotes the AN. The received signal at $U_k$, $k\in\{1,2\}$, is then given by
\begin{align}
y_k = \mathbf{h}_k^H(\mathbf{Q})\mathbf{w}_1 s_1 + \mathbf{h}_k^H(\mathbf{Q})\mathbf{w}_0 s_0 + \mathbf{h}_k^H(\mathbf{Q})\mathbf{v}z + n_k,
\end{align}
where $n_k\sim\mathcal{CN}(0,\sigma^2)$ denotes the additive white Gaussian noise (AWGN) with identical variance $\sigma^2$ at both UEs, and $(\cdot)^{H}$ is the Hermitian transpose.

At both UEs, the multicast signal $s_0$ is decoded first by treating the confidential signal and AN as interference. The corresponding SINR for decoding $s_0$ at $U_k$ is given by
\begin{align}
\gamma_{0,k}=\frac{|\mathbf{h}_k^H(\mathbf{Q})\mathbf{w}_0|^2}
{|\mathbf{h}_k^H(\mathbf{Q})\mathbf{w}_1|^2+|\mathbf{h}_k^H(\mathbf{Q})\mathbf{v}|^2+\sigma^2}.
\end{align}
To guarantee the service quality of the multicast transmission, the SINR constraint $\gamma_{0,k}\ge\gamma_{\mathrm{th}}$ is imposed.

After successfully decoding and removing the multicast signal via successive interference cancellation, $U_1$ decodes its confidential signal, and the achievable rate at $U_1$ is given by
\begin{align}
R_1=\log_2\!\left(1+\frac{|\mathbf{h}_1^H(\mathbf{Q})\mathbf{w}_1|^2}
{|\mathbf{h}_1^H(\mathbf{Q})\mathbf{v}|^2+\sigma^2}\right).
\end{align}
Similarly, $U_2$ may attempt to decode the confidential signal after removing the multicast signal, and the corresponding eavesdropping rate is given by
\begin{align}
R_{\mathrm{E}}=\log_2\!\left(1+\frac{|\mathbf{h}_2^H(\mathbf{Q})\mathbf{w}_1|^2}{|\mathbf{h}_2^H(\mathbf{Q})\mathbf{v}|^2+\sigma^2}\right).
\end{align}
Accordingly, the achievable secrecy rate at $U_1$ is defined as
\begin{align}\label{eq:Secracyrateo}
R_{\mathrm{s}}=\left[R_1-R_{\mathrm{E}}\right]^+,
\end{align}
where $[x]^+\triangleq \max\{x,0\}$.

\subsection{Problem Formulation}

We aim to maximize the secrecy rate at $U_1$ in the considered system by jointly optimizing the MA position matrix $\mathbf{Q}$, the beamforming vectors $\mathbf{w}_1$ and $\mathbf{w}_0$, and the AN vector $\mathbf{v}$, while ensuring the SINR requirements for decoding the multicast signal at both UEs. To this end, the optimization problem is formulated as

\begin{subequations}\label{eq:opt}
\begin{align}
\mathbf{P1}:\quad  \max_{\mathbf{Q},\,\mathbf{w}_1,\,\mathbf{w}_0,\,\mathbf{v}}
\quad & R_{\mathrm{s}} \label{eq:opt1} \\
\mathrm{s.t.}\quad
& \gamma_{0,k} \ge \gamma_{\mathrm{th}}, \quad k\in\{1,2\}, \label{eq:opt2} \\
& \|\mathbf{w}_1\|^2 + \|\mathbf{w}_0\|^2 + \|\mathbf{v}\|^2 \le P_{\max}, \label{eq:opt3} \\
& \mathbf{q}_m \in \mathcal{Q}, \quad \forall m, \label{eq:opt4} \\
& \|\mathbf{q}_i - \mathbf{q}_j\| \ge d_{\min}, \quad \forall i\neq j,\label{eq:opt5}
\end{align}
\end{subequations}
where \eqref{eq:opt3} denotes the total transmit power constraint at the BS, with $P_{\max}$ representing the maximum transmit power.

\section{Analytical Insights}\label{Sec:Anal}

It is important to emphasize that problem \textbf{P1} is inherently non-convex and computationally formidable. This complexity arises from the intricate interdependency among the MA position matrix $\mathbf Q$, the transmit beamforming vectors $\mathbf w_1$ and $\mathbf w_0$, and the AN vector $\mathbf v$ in both the objective function and the multicast SINR constraints in \eqref{eq:opt}, rendering the globally optimal solution intractable. Before developing an algorithm to solve \textbf{P1}, we first offer important design insights by separately considering single-MA and multi-MA cases.

\subsection{Single-MA Case}

For the single-MA case, i.e., $M=1$, the transmit beamforming vectors reduce to scalar transmit coefficients. We denote $p_1$, $p_0$, and $p_v$ as the transmit powers allocated to the confidential signal, the multicast signal, and the AN, respectively. The transmit power constraint is then given by $p_1+p_0+p_v\le P_{\max}$. Accordingly, the secrecy rate in \eqref{eq:Secracyrateo} is simplified as
\begin{align}\label{eq:Rs_pv_pf}
R_{\mathrm{s}}\!=\!
\left[\!\log_2\!\left(\!
\frac{(|h_2(\mathbf{q})|^2 p_v+\sigma^2)(|h_1(\mathbf{q})|^2(p_1+p_v)+\sigma^2)}
{(|h_1(\mathbf{q})|^2 p_v+\sigma^2)(|h_2(\mathbf{q})|^2(p_1+p_v)+\sigma^2)}\!\right)\!\right]^{+}.
\end{align}
Based on \eqref{eq:Rs_pv_pf}, the following Lemma revisits a known property of AN in single-antenna systems, adapted to the single-MA case.

\begin{Lemma}\label{Lemma:singlecase}
For the single-MA case, i.e., $M=1$, allocating the transmit power to AN does not improve the achievable secrecy rate. The optimal solution therefore satisfies $p_v^{\ast} = 0$.
\begin{IEEEproof}
The result follows from the monotonicity of the secrecy rate with respect to $p_v$ and is consistent with known results in single-antenna secure communication systems.
\end{IEEEproof}
\end{Lemma}

\textbf{Remark 1:} Lemma \ref{Lemma:singlecase} implies that, although a single MA allows channel reshaping, it still behaves as a single-antenna system and cannot enable effective AN design. In contrast, when $M \ge 2$, multiple MAs provide additional spatial DoFs for directional transmission and interference shaping, which motivates us to investigate the multi-MA case.

\subsection{Multi-MA Case}

We now extend the analysis to the general multi-MA case with $M \ge 2$.
For this case, the AN vector is expressed as $\mathbf v=\sqrt{p_v}\mathbf{\tilde{v}}$, where $p_v\ge 0$ and $\|\mathbf{\tilde{v}}\|^2=1$. The received AN power at $U_k$ is then given by
\begin{align}
|\mathbf h_k^H(\mathbf Q)\mathbf v|^2 = p_v |\mathbf h_k^H(\mathbf Q)\mathbf{\tilde{v}}|^2.
\end{align}
Under the zero AN leakage to the confidential UE, the optimal AN direction is characterized in the following proposition.

\begin{Proposition}\label{prop:AN_structure}
When $\mathbf h_1(\mathbf Q)$ and $\mathbf h_2(\mathbf Q)$ are linearly independent, under the zero AN leakage constraint $\mathbf h_1^H(\mathbf Q)\mathbf{\tilde{v}}=0$, the AN direction that maximizes the interference power at the eavesdropper is given by
\begin{align}\label{eq:proposition1}
\mathbf{\tilde{v}}^\star=\frac{\mathbf{\Psi}_{\mathbf{h}_1}\mathbf h_2(\mathbf Q)}
{\|\mathbf{\Psi}_{\mathbf{h}_1}\mathbf h_2(\mathbf Q)\|},
\end{align}
where $\mathbf{\Psi}_{\mathbf{h}_1}\!=\!\mathbf I\!-\!\frac{\mathbf h_1(\mathbf Q)\mathbf h_1^H(\mathbf Q)}{\|\mathbf h_1(\mathbf Q)\|^2}$ and $\mathbf I$ is the identity matrix.
\begin{IEEEproof}
Under the zero-AN-leakage constraint, $\mathbf{\tilde{v}}$ lies in the null space of $\mathbf h_1^H(\mathbf Q)$, i.e., $\mathbf{\tilde{v}}=\mathbf{\Psi}_{\mathbf h_1}\mathbf{\tilde{v}}$, which implies
\begin{align}
|\mathbf h_2^H(\mathbf Q)\mathbf{\tilde{v}}|^2
=
|(\mathbf{\Psi}_{\mathbf h_1}\mathbf h_2(\mathbf Q))^H\mathbf {\tilde{v}}|^2.
\end{align}
Using the Cauchy--Schwarz inequality and $\|\mathbf{\tilde{v}}\|=1$, we have
\begin{align}
|\mathbf h_2^H(\mathbf Q)\mathbf{\tilde{v}}|^2
\le
\|\mathbf{\Psi}_{\mathbf h_1}\mathbf h_2(\mathbf Q)\|^2,
\end{align}
with equality if and only if $\mathbf{\tilde{v}}$ is collinear with $\mathbf{\Psi}_{\mathbf h_1}\mathbf h_2(\mathbf Q)$. Thus, the optimal AN direction is given by \eqref{eq:proposition1}.
\end{IEEEproof}
\end{Proposition}

\textbf{Remark 2:} Proposition \ref{prop:AN_structure} elucidates that, under the zero-leakage constraint, the optimal AN is strictly confined to the null space of the confidential channel while maximizing its projection onto the eavesdropper's channel. This enables effective spatial interference shaping, which is fundamentally unavailable in the single-MA case.

To further characterize the power allocation structure, we define the AN-to-confidential power ratio as $\rho=\frac{p_v}{p_1}$, and denote $P_{\mathrm r}$ as the remaining power budget after allocating the multicast signal, i.e., $P_{\mathrm r}=P_{\max}-\|\mathbf w_0\|^2$, such that $p_1+p_v=P_{\mathrm r}$.
To obtain a tractable design, we adopt a structured beamforming strategy. Specifically, the confidential beamforming vector is chosen as $\mathbf w_1=\sqrt{p_1}\tilde{\mathbf w}_1$, where $\tilde{\mathbf w}_1 = \frac{\mathbf{h}_1(\mathbf Q)}{\|\mathbf{h}_1(\mathbf Q)\|}$, corresponding to maximum ratio transmission (MRT) towards the confidential UE. This choice maximizes the received signal power at the confidential UE and is known to be optimal for single UE transmission. Meanwhile, the AN is designed to lie in the null space of $\mathbf h_1(\mathbf Q)$ to avoid leakage. Under this structured design, the optimal AN power allocation reduces to a one-dimensional optimization over $\rho$, as characterized in the following Lemma.
\begin{Lemma}\label{lemma:rho_opt}
Under the structured beamforming design, for fixed $\mathbf Q$, $\mathbf w_0$, $\tilde{\mathbf w}_1$, and $\mathbf{\tilde{v}}$, the optimal AN-to-confidential power ratio that maximizes the secrecy rate is derived by
\begin{align}\label{eq:opt_rho}
\rho^\star=\left\{\begin{aligned}
&\left[\frac{g_1-g_2-\sqrt{\Delta}}{g_2-(1+g_3)g_1}\right]^{+},
&& \text{if } \Delta\geq0,\\
&0, && \text{otherwise},
\end{aligned}\right.
\end{align}
where $g_1=\frac{1}{\sigma^2}|\mathbf h_1^H(\mathbf Q)\tilde{\mathbf w}_1|^2 P_{\mathrm r}$, $g_2=\frac{1}{\sigma^2}|\mathbf h_2^H(\mathbf Q)\tilde{\mathbf w}_1|^2 P_{\mathrm r}$, $g_3=\frac{1}{\sigma^2}|\mathbf h_2^H(\mathbf Q)\mathbf{\tilde{v}}|^2 P_{\mathrm r}$, and $\Delta = g_1 g_2 g_3 \left(g_1+\frac{g_3-g_2}{1+g_3}\right)$.
\begin{IEEEproof}
The secrecy rate in \eqref{eq:Secracyrateo} can be written as $R_s=\frac{\phi(\rho)}{\ln 2}$, where
\begin{align}
\phi(\rho)=\ln\left(
\frac{(1+\rho+g_1)\big(1+(1+g_3)\rho\big)}
{(1+\rho)\big(1+(1+g_3)\rho+g_2\big)}
\right).
\end{align}
Since $\rho\ge 0$, the optimal solution that maximizes $\phi(\rho)$ lies either at the boundary $\rho=0$ or at an interior point. To find this optimal solution, we derive the first derivative of $\phi(\rho)$ with respect to $\rho$ as
\begin{align}\label{eq:derivative_rho}
\frac{d}{d\rho}\phi(\rho)=&\frac{1}{1+\rho+g_1}+\frac{1+g_3}{1+(1+g_3)\rho}\notag\\
&-\frac{1}{1+\rho}-\frac{1+g_3}{1+(1+g_3)\rho+g_2}.
\end{align}
For any interior optimal point, the first-order optimality condition requires $\frac{d}{d\rho}\phi(\rho)=0$. After algebraic simplification, \eqref{eq:derivative_rho} reduces to a quadratic equation, given by
\begin{align}\label{eq:firstderirho}
&(1+g_3)(g_2-(1+g_3)g_1)\rho^2
+2(1+g_3)(g_2-g_1)\rho \notag\\
&\quad
+g_2(1+g_3)+g_1(g_2g_3-1)=0.
\end{align}
Solving \eqref{eq:firstderirho} and combining it with the constraint $\rho\ge 0$ yields the optimal AN-to-confidential power ratio in \eqref{eq:opt_rho}.
\end{IEEEproof}
\end{Lemma}

\textbf{Remark 3:} 
Lemma \ref{lemma:rho_opt} reveals that, within the structured design, the AN-to-confidential power ratio is governed by relative channel gains and the effectiveness of AN projection. Specifically, when the confidential link is sufficiently stronger than the unintended link, i.e., $g_1 \gg g_2$, the optimal ratio reduces to $\rho^\star=0$, signifying that AN is unnecessary and beamforming alone is sufficient to ensure secrecy. In contrast, when AN can be effectively directed towards the unintended receiver, characterized by a larger $g_3$, allocating power to AN becomes beneficial and $\rho^\star$ increases accordingly. Furthermore, the condition $\Delta\geq0$ provides a simple criterion for AN activation. Crucially, this result transforms a potentially high-dimensional power allocation problem into a one-dimensional optimization, offering a significant reduction in computational complexity.

\section{Scheme Design}\label{Sec:Alg}

Building upon the analysis in Section \ref{Sec:Anal}, we develop a BCA-based scheme to tackle problem \textbf{P1}, as BCA is well suited for handling such non-convex problems with coupled variables via iterative block-wise optimization \cite{Razaviyayn2013,wang2026distributedoptimizationlearninggraphtransformers}. By partitioning the optimization variables into transmit variables and MA positions, the original problem is decomposed into two subproblems that can be solved iteratively. Specifically, for a given MA position matrix $\mathbf Q$, we optimize the transmit variables $(\mathbf w_1,\mathbf w_0,\mathbf v)$, and then update $\mathbf Q$ based on the optimized transmit variables. The resulting subproblems are formulated as
\begin{subequations}\label{eq:P2_alg}
\begin{align}
\mathbf{P2}:\quad
\max_{\mathbf w_1,\mathbf w_0,\mathbf v}\quad & R_{\mathrm s} \\
\mathrm{s.t.}\quad
& \gamma_{0,k}\ge\gamma_{\mathrm{th}},\quad k\in\{1,2\},\\
& \|\mathbf w_1\|^2+\|\mathbf w_0\|^2+\|\mathbf v\|^2\le P_{\max}
\end{align}
\end{subequations}
and
\begin{subequations}\label{eq:P3_alg}
\begin{align}
\mathbf{P3}:\quad
\max_{\mathbf Q}\quad & R_{\mathrm s}\\
\mathrm{s.t.}\quad
& \gamma_{0,k}\ge\gamma_{\mathrm{th}},\quad k\in\{1,2\},\\
& \mathbf q_m\in\mathcal Q,\quad \forall m,\\
& \|\mathbf q_i-\mathbf q_j\|\ge d_{\min},\quad \forall i\neq j,
\end{align}
\end{subequations}
respectively. These two subproblems are solved alternately until convergence.


\subsection{Transmit Variable Optimization with Given MA Position}

To obtain a low-complexity yet effective design, we adopt the structured transmission strategy motivated by the analysis in Section \ref{Sec:Anal}. Specifically, the AN direction is chosen as in \eqref{eq:proposition1}, which suppresses AN leakage to $U_1$ while enhancing its interference effect on $U_2$. For the multicast beamformer, we express $\mathbf w_0=\sqrt{p_0}\tilde{\mathbf w}_0$, where $\tilde{\mathbf w}_0$ denotes the normalized beam direction. To balance the reception of the multicast message at both UEs, we adopt a parameterized common beamformer \cite{Mao2022CST}, given by
\begin{align}
\tilde{\mathbf w}_0=
\frac{\delta\frac{\mathbf h_1(\mathbf Q)}{\|\mathbf h_1(\mathbf Q)\|}+(1-\delta)\frac{\mathbf h_2(\mathbf Q)}{\|\mathbf h_2(\mathbf Q)\|}}{\left\|\delta\frac{\mathbf h_1(\mathbf Q)}{\|\mathbf h_1(\mathbf Q)\|}+(1-\delta)\frac{\mathbf h_2(\mathbf Q)}{\|\mathbf h_2(\mathbf Q)\|}\right\|},
\end{align}
where $\delta\in[0,1]$ controls the tradeoff between two UEs. To maximize the secrecy rate, it is desirable to maximize the remaining power budget $P_{\mathrm r} = P_{\max}-\|\mathbf{w}_0\|^2$ for confidential signal and AN transmission. This can be achieved by minimizing the required multicast power $p_0$ while satisfying the SINR constraints. With the beam directions determined, the remaining task is to allocate transmit powers. By introducing the AN-to-confidential power ratio $\rho=\frac{p_v}{p_1}$ and leveraging Lemma \ref{lemma:rho_opt}, the optimal ratio $\rho^\star$ can be obtained in closed form. Accordingly, the confidential-signal and AN power are given by $p_1=\frac{P_{\mathrm r}}{1+\rho^\star}$ and $p_v=\frac{\rho^\star P_{\mathrm r}}{1+\rho^\star}$, respectively. Therefore, for a given $\delta$, the minimum multicast power $p_0$ required to satisfy the SINR constraints can be determined, which in turn specifies $P_{\mathrm r}$ and the corresponding transmit powers $(p_1,p_v)$. As a result, problem $\mathbf{P2}$ can be efficiently reduced to a one-dimensional search over $\delta\in[0,1]$. For each candidate $\delta$, we compute the minimum feasible $p_0$ and evaluate the corresponding secrecy rate. The optimal $\delta$ is then selected to maximize the secrecy rate.

\subsection{MA-Position Block Update for Given Transmit Variables}

For fixed $(\mathbf w_1,\mathbf w_0,\mathbf v)$, problem \textbf{P3} remains non-convex due to the nonlinear dependence of channel vectors on MA positions. To tackle this complexity, we adopt an antenna-wise local search strategy to update the MA-position block. Specifically, at each BCA iteration, the antennas are updated sequentially. When updating the $m$-th MA, the positions of all other antennas are fixed, and a local candidate set is generated around the current position $\mathbf q_m^{(t)}$, where $t$ denotes the BCA iteration index. This candidate set consists of both grid-based perturbation points and randomly generated feasible points within a local neighborhood. Any candidate violating the aperture or minimum-spacing constraints is discarded.

For each feasible candidate, the corresponding channel vectors are recomputed, and the secrecy rate is evaluated under the fixed transmit variables. The candidate yielding the highest feasible secrecy rate is selected as the updated position of the $m$-th MA. After all antennas are successively updated, an intermediate MA position matrix, denoted by $\bar{\mathbf Q}^{(t)}$, is obtained. To enhance the stability of the BCA-based scheme, a damping step is applied as $\mathbf Q^{(t)}=(1-\beta_{t-1})\mathbf Q^{(t-1)}+\beta_{t-1}\bar{\mathbf Q}^{(t)}$, where $\beta_t\in(0,1]$ is the damping factor. If the updated position matrix does not improve the secrecy rate, the local search radius is reduced according to $\Delta^{(t+1)}=\max\{\eta \Delta^{(t)},\Delta_{\min}\}$, where $\eta\in(0,1)$ is the shrinking factor and $\Delta_{\min}$ is the minimum allowable step size. This adaptive strategy enables coarse exploration in the early stage and refined local adjustment in the later stage.

\subsection{Overall Scheme and Its Complexity}

Starting from an initial MA position matrix $\mathbf Q^{(0)}$, the proposed BCA-based scheme alternately updates the transmit-variable and MA-position blocks until the improvement in the secrecy rate falls below a prescribed threshold or a maximum number of iterations is reached. Since each block-update retains the best feasible candidate, the secrecy rate is guaranteed to be non-decreasing over iterations. Moreover, as the objective is upper bounded by the finite power budget, the proposed BCA-based scheme converges to a locally optimal solution. Accordingly, we denote $N_p$ as the number of search points for {$\delta$}, $N_c$ as the number of local candidates examined for each antenna, $N_s$ as the number of inner sweeps in the MA-position update, and $I_{\mathrm{BCA}}$ as the total number of BCA iterations. The overall computational complexity is then approximately calculated as $\mathcal O\left(I_{\mathrm{BCA}}\left(N_p M+N_sN_cM^2L\right)\right)$, which is significantly lower than that of exhaustive joint optimization over continuous MA positions and transmit variables.

\section{Numerical Results}\label{Sec:Num}

In this section, we present numerical results to validate the analytical insights in Section \ref{Sec:Anal} and demonstrate the effectiveness of the proposed scheme in Section \ref{Sec:Alg}. Unless otherwise specified, the carrier frequency is set to $f_c=2.8$ GHz, corresponding to a wavelength $\lambda = c / f_c$. The BS is equipped with $M=4$ MAs confined within a square aperture region $\mathcal{Q}$ of size $6\lambda \times6\lambda$, with a minimum antenna spacing of $d_{\min}=\lambda/2$. The maximum transmit power is $P_{\max}=5$ dBm and the noise power is $\sigma^2=-104$ dBm. The distance between the BS and each UE is set to $d_k=70$ m, $k\in\{1,2\}$. The target SINR threshold for the multicast signal is set to $\gamma_{\mathrm{th}}=3$ dB. For the channel model, each UE experiences the same number of propagation paths, i.e., $L_k=L$. The complex gain of the $\ell$-th path from the BS to $U_k$ is modeled as $\alpha_{k,\ell}\sim\mathcal{CN}\left(0,\left(\frac{\lambda}{4\pi}\right)^2 \frac{d_k^{-2.5}}{L_k}\right)$, where the factor $1/L_k$ accounts for the power allocation across all paths and $2.5$ is the path-loss exponent. For comparison, we employ a fixed-antenna MRT scheme with AN \cite{Mei2017tvt} and an MA-enabled scheme without AN proposed in \cite{Shen2025wcl} as benchmark schemes.

\begin{figure}
    \centering
    \includegraphics[width=\columnwidth]{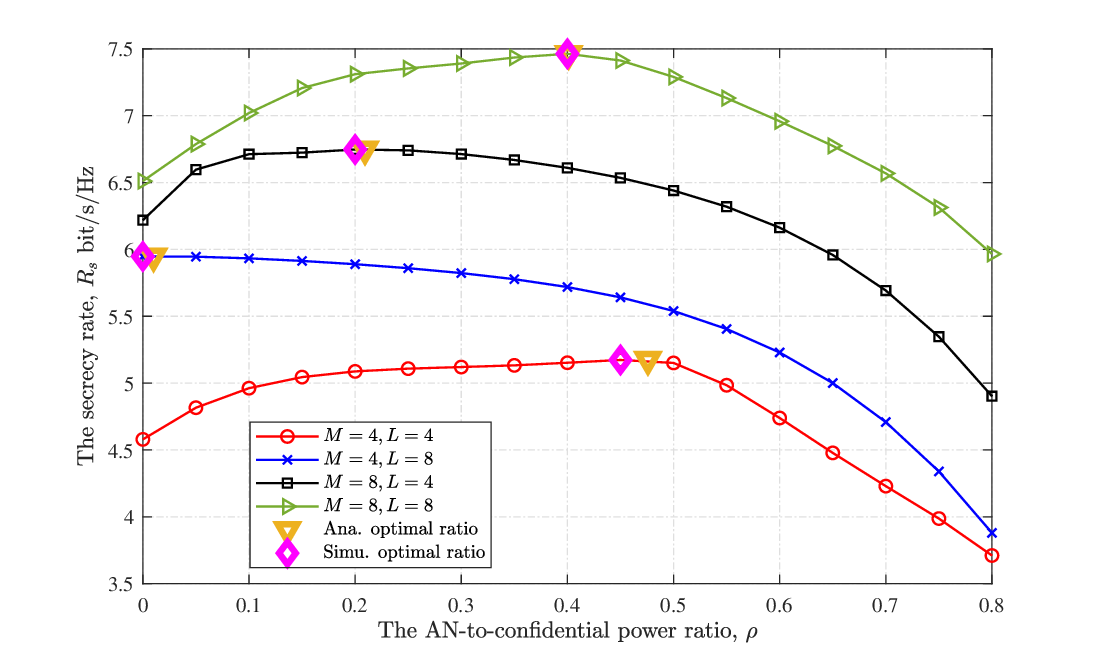}\vspace{-1em}
    \caption{The secrecy rate, $R_S$, versus the AN-to-confidential power ratio, $\rho$.}\vspace{-1em}
    \label{fig:num1}
\end{figure}

Fig. \ref{fig:num1} plots the secrecy rate, $R_S$, versus the AN-to-confidential power ratio, $\rho$. We first observe that the optimal AN-to-confidential power allocation ratio derived in Lemma \ref{lemma:rho_opt} achieves the maximum secrecy rate, which validates the accuracy of our analysis and its effectiveness in guiding power allocation design. We then observe that $R_S$ first increases and then decreases as $\rho$ increases. This observation is due to the inherent tradeoff introduced by AN. When $\rho$ is small, increasing $\rho$ enhances the AN power received at $U_2$, thereby effectively degrading the eavesdropping capability of $U_2$ and improving the secrecy rate. However, when $\rho$ becomes large, further increasing $\rho$ significantly reduces the power allocated to the confidential signal, which degrades the received signal quality at $U_1$ and thus lowers the achievable secrecy rate.

\begin{figure}[t!]
    \centering
    \includegraphics[width=\columnwidth]{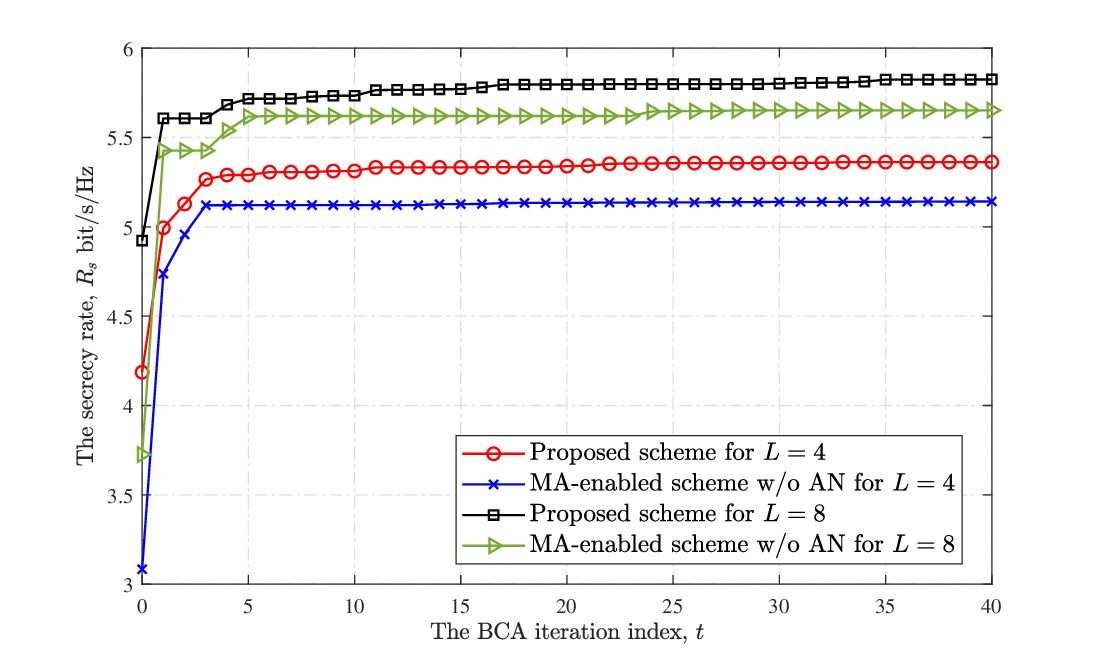}\vspace{-1em}
    \caption{The secrecy rate, $R_{s}$, versus the BCA iteration index, $t$.}\vspace{-1em}
    \label{fig:1}
\end{figure}

Fig. \ref{fig:1} plots the secrecy rate, $R_S$, versus the BCA iteration index, $t$. We first observe that the proposed scheme converges within a small number of iterations for both $L=4$ and $L=8$, indicating its favorable convergence behavior. Moreover, it consistently achieves a higher secrecy rate than the MA-enabled scheme without AN, highlighting the effectiveness of the joint design of MAs and AN. Furthermore, the secrecy rate increases monotonically over iterations, which confirms the stability of the proposed BCA-based scheme in handling the non-convex optimization problem. In addition, we observe that increasing the number of propagation paths leads to a higher converged secrecy rate, yet exerts a negligible impact on the convergence behavior. This suggests that the proposed scheme can effectively exploit richer multipath environments without incurring additional convergence overhead.

\begin{figure}[t!]
    \centering
    \includegraphics[width=\columnwidth]{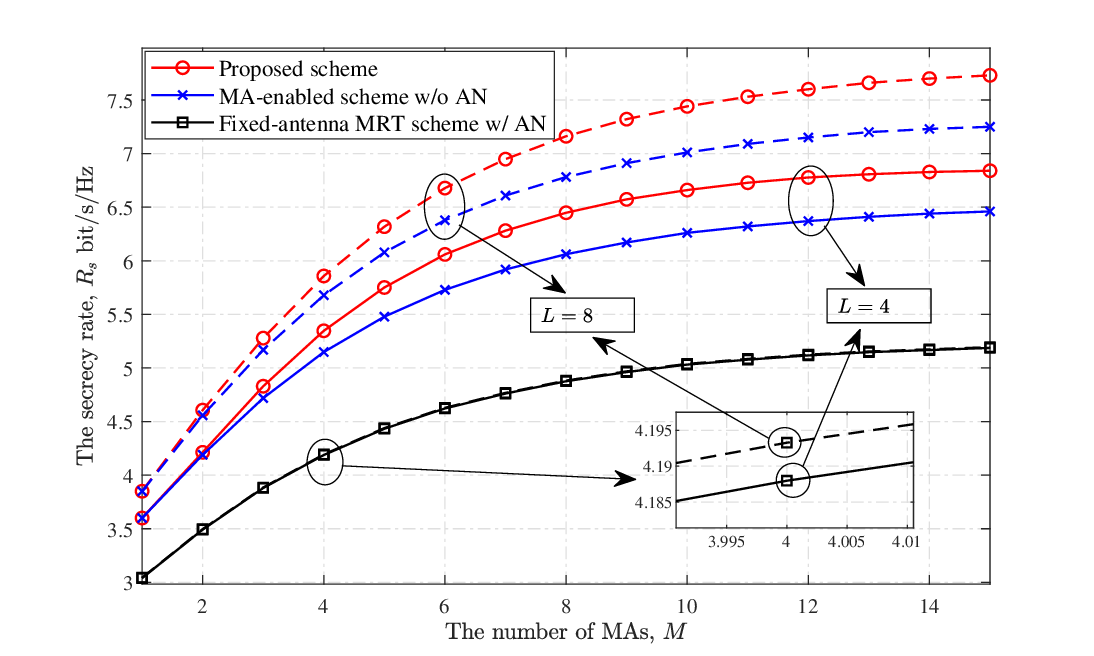}\vspace{-1em}
    \caption{The secrecy rate, $R_S$, versus the number of MAs, $M$.}\vspace{-1em}
    \label{fig:3}
\end{figure}

Fig.~\ref{fig:3} plots the secrecy rate, $R_S$, versus the number of MAs, $M$. We first observe that $R_S$ monotonically increases with $M$ for all schemes, since additional MAs provide more spatial DoFs for flexible beamforming and improved channel conditions at $U_1$. For $M=1$, incorporating AN does not improve the secrecy rate, which is consistent with Remark 1. The secrecy rate improvement from deploying more MAs is more pronounced in the small $M$ regime, but gradually diminishes for large $M$ due to the limited aperture size. Specifically, when $M$ is small, newly added MAs can effectively explore distinct spatial locations to enhance beamforming or AN design. In contrast, when $M$ grows, the aperture region becomes increasingly saturated, leaving fewer opportunities for additional MAs to achieve further gains, resulting in diminishing returns. We then observe that increasing the number of propagation paths improves the secrecy rate in MA-enabled schemes, but imposes a negligible impact on the fixed-antenna scheme. This is because richer multipath environments provide additional spatial diversity that can be effectively exploited by MAs through spatial reconfiguration, whereas such diversity cannot be effectively leveraged by fixed-antenna systems. Furthermore, the proposed scheme consistently outperforms the benchmark schemes for $M \geq 2$, validating the significant synergy between spatial reconfiguration and AN design.

\section{Conclusion}

In this paper, we investigated AN-aided MA-enabled PLSI systems and devised the joint design of MA positions and transmit variables for secrecy enhancement under multicast reliability constraints. Specifically, we characterized the AN direction under a structured transmission design and derived a closed-form expression for the AN-to-confidential power allocation ratio, which profoundly simplifies the joint design. Based on this, we developed a low-complexity BCA-based scheme that orchestrates the joint optimization of transmit variables and MA positions. Numerical results demonstrated that our proposed scheme achieves significant secrecy performance gains over benchmark schemes, while ensuring stable convergence behavior and low computational complexity, positioning MA-enabled PLSI as a cornerstone for the next-generation secure integrated service.


\bibliographystyle{IEEEtran} 
\bibliography{bibli}

\end{document}